\documentclass[pre,twocolumn,showpacs,floatfix]{revtex4}

\usepackage{amsmath}
\usepackage{amssymb}
\usepackage{graphicx}
\usepackage[mathscr]{eucal}

\begin{document}

\title{Generalized coarse graining applied to a $\phi^4$-theory:  
A model reduction-renormalization group synthesis}

\date{\today}

\author{David E. Reynolds}

\affiliation{ICES, The University of Texas at Austin,
Austin, TX 78712}

\begin{abstract}
We develop an algorithmic, system-specific
renormalization group (RG) procedure by implementing 
a systematic coarse graining procedure that 
is adapted from model reduction techniques from engineering 
control theory.  The resulting ``generalized'' RG
is a consistent generalization of the Wilsonian RG.
We  apply the generalized RG to a $\phi^4$ field theory. 
By considering nonequilibrium in addition to equilibrium observables,
we find that na\"ive power counting breaks down.  Additionally, a large 
class of short-wavelength perturbations can drive the system away from 
both the Gaussian and Wilson-Fisher fixed points.   
\end{abstract}


\maketitle



Although it is already known that the renormalization group (RG) is not 
a black box routine, the purpose of this letter is to make it more algorithmic.
  It is easy to be misled into thinking that
the  RG already is algorithmic
because its key ingredients are coarse graining 
and rescaling the system variables~\cite{goldenfeld92,shankar94}.
However, fully algorithmic implementations of the RG fail 
for large classes of problems because 
it is not possible to ignore the physics of a system and 
expect to obtain meaningful results.  
Capturing the essential physics requires isolating
the appropriate models
and the structure of perturbations and uncertainties.
Consequently, this process is system-specific.
Additionally, the scale on which the physics is observed must 
be specified.
For instance, for bosonic theories  
the long-wavelength physics is investigated,
while for fermionic system it is the physics near the Fermi surface. 
These considerations suggest that the primary obstacles to automation 
are the model identification and coarse graining processes.
In this article we present a RG procedure
that accounts for these system-specific obstacles  
and apply it to investigate the equilibrium properties of a $\phi^4$-theory.

The perspective adopted in this letter is that the obstacles mentioned above
arise from the identification of physical observables.
We use control theoretic techniques for open systems to systematically identify observables.
Specifically, consider an open system of the form
\begin{equation}
\label{open_system1}
\dot{{\bf x}} ={\bf f}({\bf x}) + {\bf u},
\end{equation}
where ${\bf x}$ and ${\bf u}$ are vectors in possibly infinite dimensional spaces.
From a field theoretic or statistical mechanical point of view,
 without ${\bf u}$, equation \eqref{open_system1} represents the ``classical'' equations of 
motion of the system.  ${\bf u}$ represents  generic driving as well as a possible noise to which 
the system is exposed.  If ${\bf u}$ is generated by a continuous stochastic process, this
 imposes a specific structure on the noise.  Theories with this restriction on ${\bf u}$
 may be mapped to a field theory via Martin-Siggia-Rose (MSR)/closed-time-path (CTP) methods~\cite{martin73,cooper01}.
In these instances, 
coarse graining is typically dictated by conserved quantities and thermodynamic considerations.
Consequently such  systems are locally coarse grained.
Suppose ${\bf u}$ is an arbitrary input (e.g.~${\bf u} \in L_2$) to the system. We consider 
the states or regions in phase space that are most accessible 
via driving to be responsible
for describing the the essential characteristics of the system.  
This is analogous to the energy landscape picture in statistical mechanics where 
fluctuations govern which states
contribute the most to the statistics of the system.
We use this control theoretic notion of the importance of states 
to specify how to coarse grain and, consequently, to generate RG equations.
The key step in generalizing the RG lies in ascertaining how to coarse grain.


%

While equation \eqref{open_system1} addresses the effects 
of perturbations to the system, it does not allow for the possibility of 
multiscale or constrained observation.
We can remedy this by considering more general open systems of the form
\begin{equation}
\label{open_system2}
\begin{array}{c}
\dot{{\bf x}} ={\bf f}({\bf x}) + B{\bf u}, \\
{\bf y} = C{\bf x},
\end{array}
\end{equation}
where ${\bf y}$ reflects that only some subspace or, more generally, subset
of phase space is directly measurable.  The operators $B$ and 
$C$ respectively specify the structure of how noise may enter the system and 
which states can be measured.
The additional structure provides a practical way to model real systems
and consider experimental constraints.  
The constraint imposed by only measuring ${\bf y}$
strongly influences the relative importance of the internal states ${\bf x}$ and hence
coarse graining.   For instance, if we measure a projected subspace of
${\bf x}$ over a finite but short 
time horizon, we can expect that those states responsible for the transient 
dynamics will be the most important.  In this case, conservation laws may
play an insignificant role in determining how to coarse grain.

In \cite{reynolds04} we proposed to coarse grain linear systems based 
on retaining the states contributing most to the response of the system to disturbances. 
Following common practice, 
we coarse grain equation \eqref{open_system2} based on its 
linearization about a particular solution with ${\bf u}=0$.  
In particular, to simplify analysis,
we only consider linearizations about equilibrium solutions.
The linearizations are generically described by  
\begin{equation}
\label{open_system4}
\begin{array}{c}
\dot{\tilde{{\bf x}}} = A\tilde{{\bf x}}+ \tilde{B}\tilde{{\bf u}}, \\
\tilde{{\bf y}} = C\tilde{{\bf x}}.
\end{array}
\end{equation}
Associated with 
equation \eqref{open_system4} are 
invariants known as Hankel singular values (HSV's)~\cite{glover84,peller03}.  
The HSV are nonnegative real numbers, $\sigma_{\max} \ge \sigma_{\kappa} \ge \sigma_{\min}$ 
that comprise the spectrum of the operator $W$.  
For systems considered over an finite time horizon, $t_f$, consider the positive operators $X$ and $Y$, 
called gramians, that are determined by the equations
\begin{eqnarray}
\label{Lyap1}
\frac{\mathrm{d}X}{\mathrm{d}t_f} = AX+XA^{\dagger}+ \tilde{B}\tilde{B}^{\dagger}; \ \ X(0)=0, \\
\label{Lyap2}
\frac{\mathrm{d}Y}{\mathrm{d}t_f} = A^{\dagger}Y+YA + C^{\dagger}C; \ \ Y(0) = 0.
\end{eqnarray}
$W^2$ may be factored as
\begin{equation}
\label{HSV}
W^2 = XY.
\end{equation}
HSV's provide a precise measure 
of the error incurred by approximating 
the effect $\tilde{\bf u}$ has on $\tilde{\bf y}$ with reduced order models.
The HSV's may be interpreted as supplying a 
measure of the importance of the internal states $\tilde{\bf x}$.  
If $W$ is invertible, it is always possible to find a coordinate system,
called {\it balanced coordinates}, such that 
$X=Y=\mathrm{diag}(\sigma_{\max},\hdots,\sigma_{\min})$~\footnote{Some subtleties arise
for infinite dimension systems, but otherwise similar results hold.}. 
When equation \eqref{open_system4} is transformed to balanced coordinates, 
the best reductions are those that project out the states corresponding 
to small HSV.  In other words, the ordering of the HSV, at least locally around
an equilibrium configuration in phase space, specifies how to coarse grain a system. 
An in depth treatment of this material may be found in \cite{reynolds04,dullerud00}.
It is also sometimes possible to ``balance'' the full nonlinear system~\cite{scherpen93}.

\begin{figure}[bp]
{\par\centering
\resizebox*{2.2in}{2.1in}
{\rotatebox{0}{\includegraphics{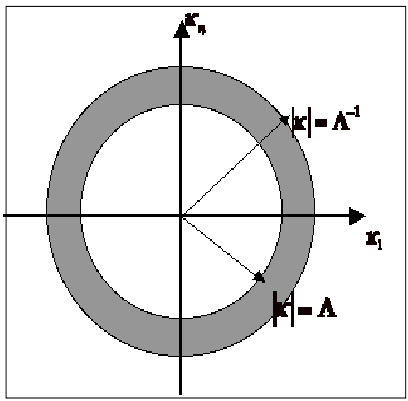}}}
\par}
\caption{ Schematic of shell in $\kappa$ space that is integrated out in the (generalized)
Wilsonian RG.}
\label{HSV_shell}
\end{figure}

%
%

The RG can easily be adapted for HSV-based coarse graining.
Operator theoretic approaches to the RG \cite{muller96,bach98} demonstrate that 
coarse graining in the Wilsonian RG is equivalent to multiplying operators or states by projection
operators \footnote{As is indicated in \cite{bach03}, we need not limit our attention to projection operators.}.
The essense of this work is to use HSV's to \emph{identify} the projection operator. 
As before, suppose that ${\bf \kappa}$ is a vector index that orders the HSV's $\sigma_{\bf \kappa}$
for equation \eqref{open_system4} from largest to smallest.
A generalized Wilsonian RG procedure is obtained by:
1) transforming the system 
to balanced coordinates (about an equilibrium solution),
\begin{equation}
\label{balancing}
\hat{\bf \phi}({\bf \kappa},t) = \int R({\bf \kappa},{\bf x}){ \bf \phi}({\bf x},t) \mathrm{d}{\bf x},
\end{equation}
so that the partition function takes the form,
\begin{equation}
\label{part_fnc1}
\begin{array}{l}
\mathscr{Z}\left[\{{\bf \phi}\}\right] = \int \mathscr{D}{\bf \phi}
\exp\left(-S\left({\bf g},\{{\bf \phi}\}\right)\right) \\ \ \ \
= \int \mathscr{D}\hat{\bf \phi}\mathscr{J}\exp\left(-S\left(\tilde{\bf g},\{\hat{\bf \phi}\}\right)\right),
\end{array}
\end{equation}
where ${\bf g}$ is the original set of coupling constants/functions, $\mathscr{J}$ is the Jacobian 
from equation \eqref{balancing}, and $\tilde{\bf g}$ is the resulting transformed set of coupling constants;
2) integrating out ${\bf \kappa}$-shells about $\sigma_{\min}$ analogously to wavevector shells;
and 3) rescaling ${\bf \kappa}$ and $\hat{\bf \phi}$ appropriately.  
An interesting but technically challenging variant of this procedure 
is to integrate out $\sigma_{\bf \kappa}$-shells instead of ${\bf \kappa}$-shells about $\sigma_{\min}$.
The remarkable feature of this variant is that $\sigma_{\bf \kappa}$
does not respect spatial dimension.  
For instance, integrating out a single  $\sigma_{\bf \kappa}$-shell
may entail integrating out an entire subspace in ${\bf x}$-space.
This seems to provide a natural way to understand how three dimensional 
systems may have regions (e.g.~affine subspaces) exhibiting one or two dimensional critical behavior.
The technical challenge lies in rescaling  $\sigma_{\bf \kappa}$.  It is not clear that rescaling 
 $\sigma_{\bf \kappa}$ will recover the full ${\bf \kappa}$-space thereby 
 generating a meaningful RG.

Before applying this procedure to the nonlinear wave equation,
we apply it to some trivial examples to 
build intuition.
%
%
We first consider the (driven) diffusion equation
\begin{equation}
\label{diffusion}
\partial_t \phi = D\nabla^2\phi + \gamma u.
\end{equation}
In this example, $B=\gamma$, $C=1$, and we take $t_f \to \infty$.
By considering a stable system over an infinite time horizon, we only need to solve the Lyapunov equations,
\begin{eqnarray}
\label{Lyap3}
AX+XA^{\dagger}+ \tilde{B}\tilde{B}^{\dagger} = 0,\\
\label{Lyap4}
A^{\dagger}Y+YA +C^{\dagger}C = 0,
\end{eqnarray}
instead of equations \eqref{Lyap1}-\eqref{Lyap2}.
By taking the Fourier transform of
equations \eqref{Lyap3}-\eqref{Lyap4} it is easy to derive
that $W$, from equation \eqref{HSV}, in balanced coordinates is given by
\begin{equation}
\label{diff_gramian}
W^{bal}_{\bf k} = \frac{|\gamma|}{2D|{\bf k}|^2}.
\end {equation}
Here the index, ${\bf \kappa}$ for the HSV's is just $|{\bf k}|$.  Thus, for the diffusion
equation, the most important states are those that correspond to small wavevector.  Thus,
local coarse graining is appropriate because the smallest observable ``fluctuations''
are due to the short-wavelength physics.
The smallest error is incurred by projecting out 
large wavevectors.
We will not complete the RG analysis here because the standard RG treatments based on 
local coarse graining are applicable.  For a field theoretic treatment it is possible to obtain 
the coarse grained action from the MSR/CTP formalism~\cite{zanella02}, however alternative formulations may 
be found in \cite{goldenfeld92, bricmont95}.


As a second example, we consider the (driven) linear wave equation
\begin{equation}
\label{linear_wave_equation1}
\begin{array}{c}
\partial_t^2 \phi = v^2\nabla^2\phi + \gamma u, \\
{\bf y} = \phi. 
\end{array}
\end{equation}
By the units of $u$, it represents a true force acting on $\phi$.  This, in addition to 
the fact that only $\phi$ is the ``measurable'' quantity, implies that we have 
isolated our attention on $\phi$-based observables.  This is choice is a very statistical
equilibrium and thermodynamic one. We have completely neglected $\pi$, the field 
conjugate to $\phi$, that represents the kinetic contributions to the system.
When posed as a set of first order equations, equation \eqref{linear_wave_equation1} becomes
\begin{equation}
\label{linear_wave_equation2}
\left[\begin{array}{c}
\partial_t \phi  \\ \partial_t \pi \\ {\bf y}  
\end{array}\right] =
\left[\begin{array}{ccc}
0 & 1 & 0 \\
v^2 \nabla^2 & 0 & \gamma \\
1 & 0 & 0 
\end{array}\right]
\left[ \begin{array}{c}
\phi \\ \pi \\ u
\end{array}\right]
\end{equation}
By smoothing out the time-cutoff at $t_f$ with a damped exponential
in the integral representation of the solution of equations \eqref{Lyap1} and \eqref{Lyap2}
the problem simplifies to solving Lyapunov equations.  This smoothing process is also 
known as exponential discounting.  With the given form of $B$ and $C$ in 
this problem, we find that the matrix of HSV's, $W$, is approximately given by
\begin{equation}
\label{lwe_gramian}
W^{bal}_{\bf k} \approx \frac{|\gamma|}{4av|{\bf k}|}\otimes I_{2\times2},
\end {equation}
where $I_{2\times 2}$ is the $2\times 2$ matrix identity, $a \sim 1/t_f$, and $\otimes$ is the dyadic 
(algebraic tensor) product. 
As with the diffusion equation, short-wavelength physics does not significantly contribute to 
the response, so locally coarse graining is appropriate. we will refrain from treating this example
in more detail because the standard RG treatment of 
the wave equation and its $\phi^4$ nonlinear generalization may be readily found 
in field theory textbooks~\cite{peskin95}.

The main point of the previous examples is to convey that the generalized Wilsonian RG  consistently 
reproduces the results of the standard RG for systems that we already intuitively know should be 
locally coarse grained.  
A great strength of the generalized RG is that it permits us to tackle less intuitive problems. 
We apply the generalized RG to determine the statistical equilibrium properties 
of a nonlinear wave equation with a particular nonequilibrium choice of observables.  
As will be seen, a surprising result is that this choice of observables forces us to 
nonlocally coarse grain.  The nonlocality of the coarse graining has very interesting implications with 
regard to the resulting induced RG flow.  In the remainder of the paper, we then consider some of 
these novel implications.

The (driven) equations of motion that we are considering are
\begin{equation}
\begin{array}{l}
\partial_t \phi = \pi + \alpha_1 u_1 \\
\partial_t \pi = \nabla^2\phi + \frac{\lambda}{3!}\phi^3 + \alpha_2 u_2 \\
{\bf y} = \left[\begin{array}{c} \beta_1\phi \\ \beta_2\pi \end{array}\right],
\end{array} 
\end{equation}
where $\phi$ and $\pi$ are real-valued fields.
The driving now includes generalized forces in addition to ``true'' forces.
By expanding around equilibrium solutions of $\nabla^2\phi=0$ we find that 
for each real-space position ${\bf x}$, 
\begin{eqnarray}
\label{BnC}
B = \left(\begin{array}{cc} \alpha_1 & 0 \\ 0 & \alpha_2 \end{array}\right) \ \textrm{and} \\ 
C = \left(\begin{array}{cc} \beta_1 & 0 \\ 0 & \beta_2 \end{array}\right).
\end{eqnarray}
This driving allows for more states in $(\phi,\pi)$-phase space
to be accessible compared to the driving in equation \eqref{linear_wave_equation2}.  
This, in combination with the form of ${\bf y}$,
ensures that both $\phi$ and $\pi$-dependent observables are being considered.
By using exponential discounting, as mentioned earlier, we find that the 
diagonal operator of HSV's is given by
\begin{equation}
\label{nlwe_gramian}
\begin{array}{c}
W_{\bf k} \approx \frac{1}{4a}\left[(\alpha_2^2|{\bf k}|^{-1}+\alpha_1^2|{\bf k}|)\right. \\
\left.\times(\beta_1^2|{\bf k}|^{-1}+\beta_2^2|{\bf k}|)\right]^{1/2}\otimes I_{2\times2}.
\end{array}
\end {equation}
$W_{\bf k}$ does not have the HSV's ordered from largest to smallest, so it is 
not expressed in balanced coordinates.  It is immediately apparent that the HSV's 
are large for both large and small magnitude wavevector.  
A heuristic explanation for this strange result is that for large wavevector,
$\pi$ is a pathologically ``fast'' variable.  However, by driving $\pi$ with $u_1$
over all wavevector, this permits the fast resonances to be excited at large wavevector.  
The pathological nature of $\pi$ as an observable
is analogous to the pathological nature of considering $\dot{\xi}$ an observable where
$\xi$ satisfies a Langevin equation~\footnote{$\dot{\xi}$ is pathologically fast compared to $\xi$}. 
For this reason, $\pi$ is a nonequilibrium observable.
Furthermore, because both the small and large wavelength 
physics contributes strongly to the response of the system, local coarse graining cannot be correct.
The appropriate coarse graining is nonlocal.
  
In the case where $\phi$ and $\pi$ are treated on equal footing as observables, which in general may 
not be the case, $\alpha_1=\alpha_2=\alpha$ and $\beta_1=\beta_2=\beta$.  In the remainder, we treat 
this particular case.  Furthermore, without loss of generality, we set $\alpha = \beta = 1$.
In this case, equation \eqref{nlwe_gramian} indicates that 
 the $|{\bf k}|=1$ states are the least important.  Implementing the second step of 
the procedure for generalized RG involves integrating out ${\bf k}$-shells {\it away} from the
$|{\bf k}|=1$ surface.  Rather than transform the system into the balanced ${\bf \kappa}$-coordinates,
out of convenience, we coarse grain the system in ${\bf k}$-space.  

The action for this system, without driving, in Fourier space is given by 
\begin{equation}
\label{action}
\begin{array}{l}
S({\bf g},\{\hat{\phi}\}) =  \frac{1}{(2\pi)^{D}}\int \mathrm{d}{\bf k} |{\bf k}|^2\hat{\phi({\bf k})}^2  +  \\
  \frac{\lambda}{4!} \int \prod_{n=1}^4 \frac{\mathrm{d}{\bf k}_n}{(2\pi)^D}  
\delta\left(\sum_{j=1}^4 {\bf k}_j\right) 
\hat{\phi}({\bf k}_1)\hat{\phi}({\bf k}_2)\hat{\phi}({\bf k}_3)\hat{\phi}({\bf k}_4), 
\end{array}
\end{equation}
where $D$ is the spatial dimension of the system since
we are only considering the statistical equilibrium properties of the system.  
If we wished to do a full nonequilibrium treatment, 
we should coarse grain the MSR/CTP-action for the system.  
In order to coarse grain, we let $\hat{\phi} = \hat{\phi}_< + \hat{\phi}_m + \hat{\phi}_>$ where 
$\hat{\phi}_<$ is only nonzero for $|{\bf k}| \le \Lambda$,
$\hat{\phi}_m$ is only nonzero for $\Lambda < |{\bf k}| < \Lambda^{-1}$, 
and $\hat{\phi}_>$ is only nonzero for $\ |{\bf k}| \ge \Lambda^{-1}$, where $\Lambda<1$.
With this decomposition, the path integral measure factors as 
$\mathscr{D}\hat{\phi}=\mathscr{D}\hat{\phi}_<\mathscr{D}\hat{\phi}_m\mathscr{D}\hat{\phi}_>$.
The RG equations are then induced by integrating out $\hat{\phi}_m$ and then rescaling 
the wavevectors and fields.  For this problem, the rescaling procedure requires that 
\begin{equation}
\label{rescale}
\begin{array}{c}
\hat{\phi}_<({\bf k}) = Z_<\varphi_1(\Lambda^{-1}{\bf k}),  \\
\hat{\phi}_>({\bf k}) = Z_>\varphi_2(\Lambda{\bf k}),
\end{array}
\end{equation}
and ${\bf p} = \Lambda^{-1}{\bf k}$ for $|{\bf k}| \le \Lambda$ and
 ${\bf p} = \Lambda{\bf k}$ for $|{\bf k}| \ge \Lambda^{-1}$. Na\"ive power
counting breaks down as a direct result of rescaling in the two disjoint wavevector 
regimes.

Although we start with a theory where ${\bf g} = (1,\lambda,0,\hdots)$ we can 
expect that the RG transformations may generate new nonlinear terms 
and that the coupling constants may become coupling functions.  
In fact, ${\bf g}$ flows towards having an 
infinite number of nontrivial components.  
In particular, the coupling constant $\lambda$ becomes a coupling function,
$\lambda({\bf p_1},{\bf p_2},{\bf p_3},{\bf p_4})$, that may be decomposed 
into the coupling functions $\{\lambda_{(i,4-i)}\}_{i=1}^4$.  Here the 
notation indicates that $\lambda_{(i,j)}$ has $i$ wavevectors with $|{\bf p}_n|<1$
and $j$ wavevectors with $|{\bf p}_n|>1$.
If we let $\Lambda = e^{-dl}$, then to first loop order the RG equations for 
$\lambda_{(i,j)}({\bf p_1},{\bf p_2},{\bf p_3},{\bf p_4})$ are
\begin{equation}
\label{RG_eqns1}
\begin{array}{l}
\partial_l\lambda_{(i,j)} =  \alpha_{i,j}\lambda_{(i,j)}  \\  + 
4^{-1}\lambda_{(i,j)}^2J({\bf p_1},{\bf p_2},{\bf p_3},{\bf p_4}) + \mathscr{O}(\lambda^3), 
\end{array}
\end{equation}
where $\alpha_{0,4} = D-4$, $\alpha_{4,0} = 4-D$, $\alpha_{2,2} = -D$, $\alpha_{3,1} = -2(D-1)$, 
and $\alpha_{1,3} = -2$. Also, $J({\bf p_1},{\bf p_2},{\bf p_3},{\bf p_4}) =$ 
$K({\bf p_1}+{\bf p_2})+K({\bf p_1}+{\bf p_3})+\hdots+K({\bf p_3}+{\bf p_4})$, where $K({\bf q})$ 
comes from the one-loop contribution to the 4-point vertex and is given by
\begin{equation}
\label{loop_term}
\begin{array}{c}
K({\bf q}) = \frac{2}{\Lambda-\Lambda^{-1}}\int_{\Lambda < |{\bf k}| < \Lambda^{-1}}
\frac{\mathrm{d}{\bf Q}}{(2\pi)^D}\frac{1}{{\bf Q}^2({\bf Q}+{\bf q})^2} \\
= \frac{2^{D-1}S_{D-1}}{(2\pi)^{D}\left(1+|{\bf q}|\right)^{2}}B\left(\frac{D-1}{2},\frac{D-1}{2}\right) \\
\times {}_2F_1\left(D-1, \frac{D-1}{2}; 1; \frac{4|{\bf q}|}{(1+|{\bf q}|)^2}\right),
\end{array}
\end{equation}
where $B(x,y)$ is the Euler beta function and ${}_2F_1$ is the hypergeometric function.
Although we do not start with a ``mass'' term in the action (i.e. $m^2\phi^2$), such a term
is generated by the RG flow.  If we denote the mass terms for $|{\bf q}|<1$  and  $|{\bf q}|>1$ 
respectively by $m^2_<({\bf q})$ and $m^2_>({\bf q})$, then the associated RG equations for these 
terms are given by
\begin{eqnarray}
\label{RG_eqns2a}
\partial_lm^2_< = 2m^2_< + \frac{S_D\lambda}{2(2\pi)^{D}}+\mathscr{O}({\lambda^2}) \\
\label{RG_eqns2b}
\partial_lm^2_> = -2m^2_> + \frac{S_D\lambda}{2(2\pi)^{D}}+\mathscr{O}({\lambda^2})
\end{eqnarray}

The first thing to notice in equation \eqref{RG_eqns1} is that the contribution from tree level,
the linear term, indicates that the coupling functions  involving a mixing of wavevectors 
(i.e.~ $i,j\ne 0 $) are irrelevant.  This is the first piece of evidence that 
the theory decouples at small and large wavelength.  The second piece of evidence for this is that
$K({\bf q})$ in equation \eqref{loop_term} diverges as $|{\bf q}|\to 1$.  Due to $K({\bf q})$-dependence $\lambda$
acquires through $J({\bf p_1},{\bf p_2},{\bf p_3},{\bf p_4})$, this divergence seems to
act like a barrier in the effective coarse grained theory to prevent the large and small wavevector 
physics from mixing.  As alluded to earlier, this decoupling of the small and large 
wavelength physics is a manifestation of independence of $\phi$ and $\pi$ as physical observables.
Rather than being general, we consider this decoupling to be special because we are only considering 
a real, scalar $\phi^4$-theory.  Without driving, this model lacks continuous symmetries to be broken.
In a more general model, the existence of such symmetries and the associated set of gauge 
transformations would provide means of coupling $\phi$ and $\pi$. 

It is not possible to ignore the wavevector dependence that $\lambda$ acquires because
they are relevant for $|{\bf p}_i|>1$.  Specifically, higher derivative perturbations, 
$p^n\hat{\phi}^2, n>2$ and $p^n\hat{\phi}^m, n>0, m \ge 4$, 
in addition to higher order nonlinearities, $\hat{\phi}^n, n>4$, become relevant when $|{\bf p}_i|>1$.  
However, the couplings at small wavevector, $|{\bf p}_i|>1$, obey the standard RG 
equations obtained by local coarse graining. 
Physically this means that just as long as the system
is not exposed to short-wavelength  perturbations, the long-wavelength, ``thermodynamic'' physics
will remain robustly observable.  If the system is perturbed to its large-wavevector regime, then 
it will flow to a short-wavelength fixed point instead of the more familiar Gaussian and Wilson-Fisher
fixed points.  This reflects that the dynamically faster short-wavelength 
perturbations are able to excite the conjugate field $\pi$, thereby driving the system away from its standard 
statistical equilibrium.  Were the conjugate field not accessible to the ``noise'', $\alpha_1=0$, or not an 
observable, $\beta_2=0$, this phenomena would not have occurred.

\begin{table}
\begin{center}
\begin{tabular}{|c|ccc|ccc|}
\hline
& $|p|<1$ & & & $|p|>1$ & & \\
coarse &$p^n\hat{\phi}^2,$ & $p^n\hat{\phi}^m,n>0,$ &  $\hat{\phi}^n,$& 
$p^n\hat{\phi}^2,$ & $p^n\hat{\phi}^m,n>0,$ &  $\hat{\phi}^n,$\\ 
graining&$n>2$& $m \ge 4$ & $n>4$ &$n>2$& $m \ge 4$ & $n>4$ \\
\hline
local & no & no & no & no & no & no \\
nonlocal & no & no & no & yes & yes & yes \\ 
\hline
\end{tabular}
\end{center}
\caption{Relevance of perturbations}
\label{relevance_table}
\end{table}

In this letter we have presented a new RG procedure and have applied it to 
a $\phi^4$ toy model.  This procedure provides, to the author's knowledge, the first systematic means to identify 
the RG projection operator.  
When both equilibrium and nonequilibrium observables are chosen, 
this RG procedure predicts that na\"ive power counting breaks down and that terms 
that are ordinarily irrelevant become relevant at large wavevector.
Table \ref{relevance_table} summarizes these results for the $\phi^4$-theory.
The generalized Wilsonian RG developed here is applicable to nonequilibrium and 
heterogeneous systems, finite or infinite dimensional systems,
 and systems with various perturbations and uncertainties.
Although the RG is still formally an uncontrolled approximation, the coarse graining is chosen
such that the effective, coarsened system is close to the original one.
Despite the versatility of this method, it is often difficult to analytically determine 
the balancing transformations.  However, since there are very efficient numerical algorithms for 
finding balanced coordinates, this generalized RG remains a numerically useful and practical algorithm.

\acknowledgments
This work was supported by NSF Grant No.~DMR-9813752.  Special thanks are due to Jean Carlson for 
her valuable comments.


\end{document}